\documentstyle[12pt,cmb,epsf,rotate]{article}

\def\bt{\begin{tabbing}}
\def\et{\end{tabbing}}
\def\beq#1{\begin{equation}\label{#1}}
\def\eeq{\end{equation}}

\def\ie{{\frenchspacing\it i.e.}}

\def\rms{{\frenchspacing r.m.s.}}

\def\kt{k_\perp}
\def\FWHM{{\rm FWHM}}

\def\spose#1{\hbox to 0pt{#1\hss}}
\def\simlt{\mathrel{\spose{\lower 3pt\hbox{$\mathchar"218$}}
     \raise 2.0pt\hbox{$\mathchar"13C$}}}
\def\simgt{\mathrel{\spose{\lower 3pt\hbox{$\mathchar"218$}}
     \raise 2.0pt\hbox{$\mathchar"13E$}}}      
\begin{document}

\heading
  {\bf CONSTRAINING TOPOLOGY WITH THE CMB}   

\author{Ang\'{e}lica de Oliveira-Costa$^{1}$, George F. Smoot$^{2}$ \& 
 Alexei A. Starobinsky$^{3}$}
{$^1$Max-Planck-Institut f\"ur Astrophysik, Garching, Germany.} 
{$^2$Lawrence Berkeley National Laboratory, Berkeley, USA.}
{$^3$Russian Academy of Sciences, Moscow, Russia.}

%\hbox{\,}
%\vskip-4.83cm
%\epsfxsize=3.7truecm\centerline{\hskip-1mm\epsfbox{fig1.ps}}
%\vskip0.2cm

\begin{abstract}{\baselineskip 0.4cm} 
We present a new data analysis method to study rectangular $T^3$ ``small 
universes" with one or two of its dimensions significantly smaller 
than the present horizon (which we refer to as $T^1$- and $T^2$-models, 
respectively). We find that the 4 year $COBE$/DMR data set a lower limit 
on the smallest cell size for $T^1$- and $T^2$-models of 3000$h^{-1}$Mpc, 
at 95\% confidence, for a scale invariant power spectrum ($n$=1). 
\end{abstract}

\section{Introduction}

\tab
In the past few years, mainly after the discovery of CMB anisotropies by 
$COBE$/DMR \cite{Smoot92}, the study of the topology of the universe has 
become an important problem for cosmologists and some hypotheses, such 
as the ``small universe" model \cite{Ellis86}, have received considerable 
attention. 
From the theoretical point of view, it is possible to have quantum creation 
of the universe with a multiply-connected topology \cite{ZelStar84}. 
From the observational side, this model has been used to explain the 
``observed" periodicity in the distributions of quasars \cite{Fang85} 
and galaxies \cite{Broad90}.

Almost all work on ``small universes" has been limited to the case 
where the spatial sections form a rectangular basic cell with sides 
$L_{x}, L_{y}, L_{z}$ and with opposite faces topologically connected, 
a topology known as toroidal. 
The three-dimensional cubic torus $T^3$ is the simplest model among all 
possible multiply-connected topologies, in which all three sides have the 
same size $L \equiv L_{x} = L_{y} = L_{z}$. 
In spite of the fact that cubic $T^3$-model has been ruled out by $COBE$ 
results \cite{dOCS95, Jing94, Sokolov, Star93, Stevens}, the possibility 
that we live in a universe with a more anisotropic topology, such as a 
rectangular torus $T^3$, is an open option that has not been ruled 
out yet.
For instance, if the toroidal model is not a cube, but a rectangle 
with sides $L_{x} \neq L_{y} \neq L_{z}$ and with one or two of its 
dimensions significantly smaller than the horizon $R_H$ 
($\equiv 2cH_{0}^{-1}$), this small rectangular universe cannot be 
completely excluded by any of the previous analyses: constraints 
from the DMR data merely require that at least one of 
the sides of the cell be larger than $R_H$. 

\begin{figure}
 \vspace{-0.3cm}
 \centerline{\rotate[r]{\vbox{\epsfxsize=9.5cm\epsfbox{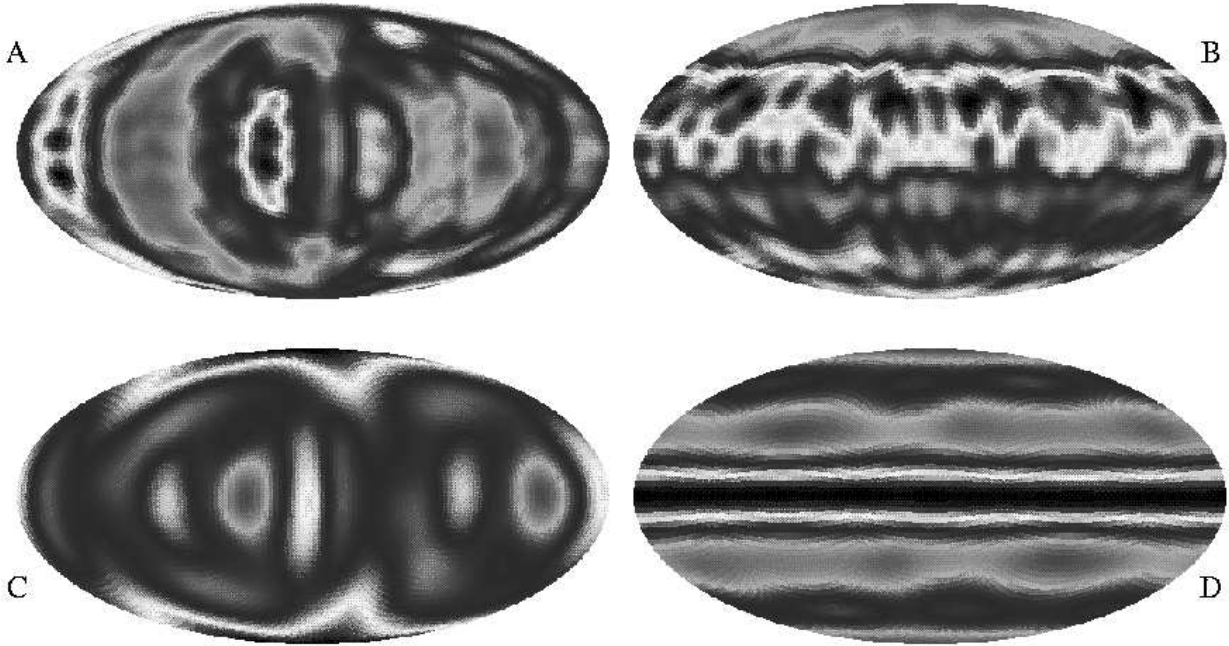}}}}
 \vspace{-0.1cm}
 \caption{\small{
Simulated sky maps for the $T^1$- and $T^2$-models and their $S$-maps.
(A) $T^1$-model with dimensions ($L_x,L_y,L_z$) = $R_H$(3,3,0.3);
(B) $T^2$-model with dimensions ($L_x,L_y,L_z$) = $R_H$(0.3,0.3,3); 
(C) $S$-map of the $T^1$-model shown in ($A$); 
(D) $S$-map of the $T^2$-model shown in ($B$).
Both  models are plotted in galactic coordinates and have a scale invariant
power spectrum ($n$=1). 
}}
 \label{fig:fourmap}
 \end{figure}

As pointed out by \cite{Fang93} and \cite{Star93}, if the rectangular 
$T^3$-universe has one of the cell sizes smaller than the horizon 
and the other two cell sizes are of the order of or larger than the horizon 
(for instance, $L_z \ll R_H$ and $L_x,L_y \simgt R_H$), 
the values of $\delta T/T$ 
are almost independent of the $z$-coordinate, \ie, the large scale CMB 
pattern shows the existence of a symmetry plane 
formed by the $x$ and $y$-axes; 
and if it has two cell sizes smaller than the horizon and the third cell size 
is of the order of or larger than the horizon (for instance, $L_x,L_y \ll R_H$ 
and $L_z \simgt R_H$), the temperatures $\delta T/T$ 
are aproximately independent 
of both $x$ and $y$, \ie, the CMB pattern shows the 
existence of a symmetry axis:
values of $\delta T/T$ are almost constant along rings 
around the $z$-axis.
We call the former case a $T^1$-model because the spatial topology of the 
universe becomes just $T^1$ in the limit $L_x,L_y \to \infty$ with $L_z$ being
fixed. The later case is denoted a $T^2$-model for the same reason (the 
corresponding limit is $L_z \to \infty$ with $L_x,L_y$ being fixed). 
See Figures~\ref{fig:fourmap}A (upper left) and~\ref{fig:fourmap}B 
(upper right). 

Our goal is to show that the $COBE$/DMR maps have the ability to test 
and rule out $T^1$- and $T^2$-models. 
We use a new approach to study these models in which we constrain their 
sizes by looking for the symmetries that they would produce in the CMB,
obtaining strong constraints from the 4 year $COBE$/DMR data.

\section{The method}

\tab
The analysis of $T^1$- and $T^2$-models is not an easy task, since there are
infinitely many combinations of different cell sizes and cell orientations. 
In order to study these models, we choose a statistic in which 
we calculate the function $S(\hat{\bf n}_{i})$ defined by \cite{dOCSS96}
\beq{Smin_plane} S(\hat{\bf n}_{i}) \equiv \frac{1}{N_{pix}} 
\sum_{j=1}^{N_{pix}} \frac{ [ \frac{\delta T}{T}(\hat{\bf n}_j) - 
\frac{\delta T}{T}(\hat{\bf n}_{ij}) ]^{2}}{ \sigma(\hat{\bf n}_j)^{2} + 
\sigma(\hat{\bf n}_{ij})^{2}}, \eeq 
where $N_{pix}$ is the number of pixels that remain in the map after the Galaxy
cut has taken place, $\hat{\bf n}_{ij}$ denotes the reflection of 
$\hat{\bf n}_{j}$ in the plane whose normal is $\hat{\bf n}_{i}$, \ie, 
\beq{normal} \hat{\bf n}_{ij} = \hat{\bf n}_{j} - 2 (\hat{\bf n}_{i} \cdot
\hat{\bf n}_{j}) \hat{\bf n}_{i}, \eeq
and $\sigma(\hat{\bf n}_{j})$ and $\sigma(\hat{\bf n}_{ij})$ are the {\rms} 
errors associated with the pixels in the directions $\hat{\bf n}_{j}$ and 
$\hat{\bf n}_{ij}$. 
$S(\hat{\bf n}_{i})$ is a measure of how much reflection symmetry there is in 
the mirror plane perpendicular to $\hat{\bf n}_{i}$. The more perfect the 
symmetry is, the smaller $S(\hat{\bf n}_{i})$ will be. 
When we calculate $S(\hat{\bf n}_{i})$ for all 6144 pixels at the positions 
$\hat{\bf n}_{i}$, we obtain a sky map that we refer to as an $S$-map. This 
sky map is a useful visualization tool and gives intuitive understanding of
how the statistic  $S(\hat{\bf n}_{i})$ works.  

In order to better understand $S(\hat{\bf n}_{i})$, we first consider the
simple model of a $T^1$-universe with $L_z \ll R_H$.
For this specific model, the values of $\delta T/T$ are almost independent 
of the $z$-coordinate, so we have almost perfect mirror symmetry about the 
$xy$-plane or, in spherical coordinates, $\delta T/T (\theta,\phi) \approx 
\delta T/T (\pi-\theta,\phi)$. 
        %\ie, $S(\hat{\bf z}) \approx S(- \hat{\bf z})$. 
When $\hat{\bf n}_{i}$ points in the direction of the smallest cell size (\ie,
in the $z$-direction), we have $S(\hat{\bf n}_{i}) \approx$ 1; otherwise, 
$S(\hat{\bf n}_{i}) >$ 1. 
An $S$-map for a $T^1$-model ($L_x,L_y,L_z$) = $R_H$(3,3,0.3) can be seen in 
Figure~\ref{fig:fourmap}C (lower left). Notice in this plot 
that the direction in 
which the cell is smallest can be easily identified by two ``dark spots" at 
$\hat{\bf n}_{i} \approx \hat{\bf z}$ and $\hat{\bf n}_{i} \approx - \hat{\bf 
z}$. For $T^2$-models, the only difference will be that in the place 
of the two 
``dark spots", we have a ``dark ring" structure in the plane formed
by the two small directions. See Figure~\ref{fig:fourmap}D (lower right), an 
$S$-map of the $T^2$-model ($L_x,L_y,L_z$) = $R_H$(0.3,0.3,3).

From the two $S$-maps, we can infer two important properties: first, 
the direction in which the $S$-map takes its minimum value, denoted 
$S_{\circ}$, is the direction in which the universe is small. For a large
universe such as $L_x, L_y, L_z \gg R_H$, 
the $S_{\circ}$-directions obtained from different 
realizations are randomly distributed in the sky. 
Secondly, the distribution of $S_{\circ}$-values changes with the cell size, 
\ie, as the universe becomes smaller, the values of $S_{\circ}$ decrease.
From the definition of the $S$-map, it is easy to see that the value of 
$S_{\circ}$ is independent of the cell orientation. In other words, 
if we rotate
the cell, we will just be rotating the $S$-map, leaving its minimum value 
$S_{\circ}$ unchanged.

From here on, we will present our results in terms of the cell sizes $R_x$, 
$R_y$ and $R_z$, usually sorted as $R_x \le R_y \le R_z$ and defined as 
$R_x \equiv L_x/R_H$, $R_y \equiv L_y/R_H$ and $R_z \equiv 
L_z/R_H$. We remind the reader that the results are identical for all six 
permutations of $R_x$, $R_y$ and $R_z$.

\section{Data Analysis}

 \begin{figure}
 \vspace{-3.3cm}
 \centerline{\rotate[l]{\vbox{\epsfxsize=12.8cm\epsfbox{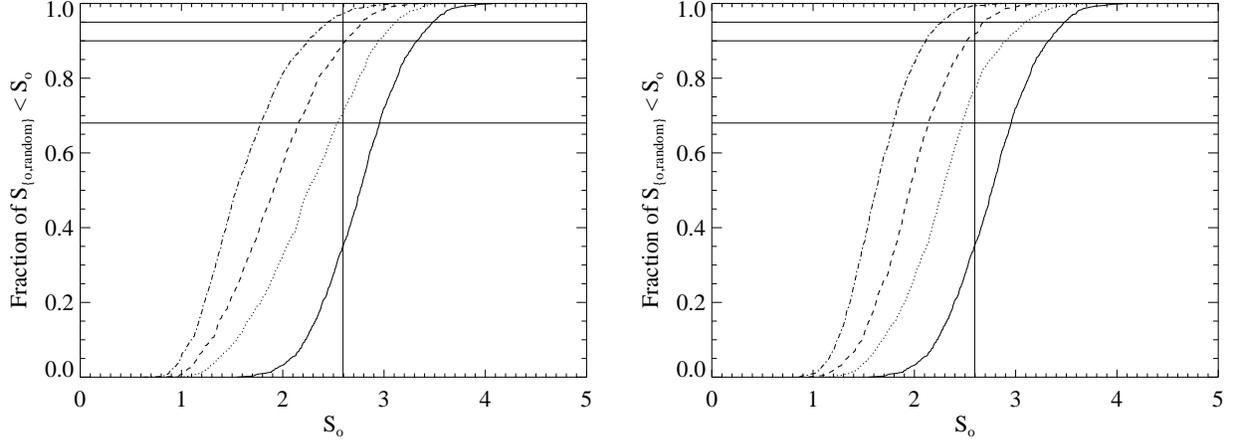}}}}
 \vspace{-3.2cm}
 \caption{\small{
Cumulative probability distribution of $S_{\circ}$ for $T^1$- and 
$T^2$-models obtained from Monte Carlo simulations. 
 A, left plot: Simulations for $T^2$-universes with dimensions 
($R_x,R_y,R_z$) = (0.5,0.5,3) or dot-dashed line, 
(0.6,0.6,3) or dashed line, and 
(0.7,0.7,3) or dotted line. 
 B, right plot: Simulations for $T^1$-universes with dimensions 
($R_x,R_y,R_z$) = 
(0.5,3,3) or dot-dashed line, 
(0.6,3,3) or dashed line, and 
(0.7,3,3) or dotted line. 
In both pictures the model ($R_x,R_y, R_z$) = (3,3,3) is represented
by a solid line, $S_{\circ}^{DMR}$ = 2.59 (vertical straight line) 
and the horizontal solid lines indicate the confidence levels of 
95\%, 90\% and 68\% (from top to bottom).
}} 
 \label{fig:histograms}
 \end{figure}

\tab
If the density fluctuations are adiabatic and the Universe is spatially
flat, the Sachs-Wolfe fluctuations in the CMB are given by 
\cite{Peebles82}
\beq{Sachs_Wolfe} \frac{\delta T}{T} (\theta,\phi) = - \frac{1}{2} 
\frac{H_{0}^{2}}{c^{2}} \sum_{\bf k} \frac{\delta_{\bf k}}{{k}^{2}} 
e^{i{\bf k} \cdot {\bf r}}, \eeq
where {\bf r} is the vector with length 
$R_H \equiv 2c H_{0}^{-1}$ that points in the direction of observation 
($\theta,\phi$), $H_{0}$ is the Hubble constant (written here as 
$100h$~km s$^{-1}$ Mpc$^{-1}$) and $\delta_{\bf k}$ is the density fluctuation 
in Fourier space, with the sum taken over all wave numbers {\bf k}. 

In a Euclidean topology the universe is isotropic, and the sum in 
(\ref{Sachs_Wolfe}) is normally replaced by an integral. 
However, in a toroidal universe this is not the case. In this model, only wave 
numbers that are harmonics of the cell size are allowed. As a result, we have 
a discrete {\bf k} spectrum \cite{Fang87, Zel73}
\beq{kcut} {\bf k} = \frac{2 \pi}{R_H} \left(  \frac{p_x}{R_x}, 
\frac{p_y}{R_y},\frac{p_z}{R_z} \right),  \eeq
where $p_x$, $p_y$ and $p_z$ are integers.   

From equation (\ref{Sachs_Wolfe}), we can construct simulated skies by
calculating \cite{dOCSS96} 
\beq{new_SW} \frac{\delta T}{T} (\theta,\phi) \propto
\sum_{p_x, p_y, p_z} \left[ g_1 \cos(2 \pi \gamma) + g_2 \sin(2 \pi \gamma) 
\right] \alpha^{\frac{n-4}{4}} e^{-(R_H \Theta \kt)^{2}/2} + 
n (\theta,\phi), \eeq
where $g_{1}$ and $g_{2}$ are two independent Gaussian random variables with 
zero mean and unit variance,                  
$\gamma = \left( \frac{p_x}{R_x} x + \frac{p_y}{R_y} y + \frac{p_z}{R_z} 
z \right)$,  
$\alpha \equiv \left( \frac{p_x}{R_x} \right)^2 + \left( \frac{p_y}{R_y} 
\right)^2 + \left( \frac{p_z}{R_z} \right)^2 \propto k^2$ and 
$n$ is the spectral index of the scalar perturbations. 
The term $e^{-(R_H \Theta \kt)^{2}/2}$ represents the 
experimental beam function,
where $\kt$ is the length of the ${\bf k}$-component perpendicular to the 
line of sight and $\Theta$ is the width of the Gaussian beam given by
$\Theta = \FWHM / \sqrt{8 \ln{2}} \approx 0.43~\FWHM$, where FWHM is the 
full width of the beam at its half maximum. Finally, we model the noise 
$n(\theta,\phi)$ at each pixel $i$ as independent Gaussian random 
variables with mean $\langle n_i \rangle =$0 and variance
$\langle n_in_j \rangle = \sigma_{ij} \delta_{ij}$ 
\cite{Line94}.

We generate our simulated skies as standard DMR maps with 6144 pixels for 
$n$=1, with a Galaxy cut of 20$^{\circ}$, FWHM = 7$^{\circ}$, the monopole and 
dipole removed, add noise and normalize to $\sigma_{7^{\circ}} = 
34.98\mu$K (the {\rms} value at 7$^{\circ}$ extracted from our DMR map, a 
4 year combined 53 plus 90 GHz map with monopole and dipole removed). 
Fixing a cell size, we construct a simulated sky according to (\ref{new_SW}), 
we smooth this once more by 7$^{\circ}$ and use the statistic defined
in (\ref{Smin_plane}) to obtain an $S$-map from which we extract its minimum 
value $S_{\circ}$. 
Repeating this procedure 1000 times, we obtain the probability distribution 
of $S_{\circ}$ for that fixed cell size. When we repeat this same procedure 
for different cell sizes, we are able to construct plots as shown in 
Figure~\ref{fig:histograms}. 

In Figure~\ref{fig:histograms}A (left plot), we show the cumulative 
probability 
distribution of $S_{\circ}$ obtained from the Monte Carlo simulations for the 
cell sizes ($R_x,R_y,R_z$) = (0.5,0.5,3), (0.6,0.6,3), (0.7,0.7,3) and
(3,3,3). 
The horizontal lines indicate the confidence levels of 95\%, 90\% and 
68\% (from top to bottom). 
Comparing these curves with the value $S_{\circ}^{DMR}$ = 2.59 (represented in 
the plot by the vertical straight line), where $S_{\circ}^{DMR}$ is the 
$S_{\circ}$ value extracted from our data set, we conclude that $T^2$-models 
with smallest cell sizes $R_x,R_y\simlt$0.5 can be ruled out at 95\% 
confidence.
As the second cell size $R_y$ is increased, the curves shift to the left of 
the $T^2$-models and we can rule out $T^1$-models for 
$R_x\simlt$0.5 at a similar confidence level, see Figure~\ref{fig:histograms}B 
(right plot). In this plot, we show the cumulative probability distribution of 
$S_{\circ}$ obtained from Monte Carlo simulations for the cell sizes 
($R_x,R_y,R_z$) = (0.5,3,3), (0.6,3,3), (0.7,3,3) and (3,3,3). 

\begin{figure}
 \vspace{-1cm}
 \centerline{\rotate[l]{\vbox{\epsfxsize=10cm\epsfbox{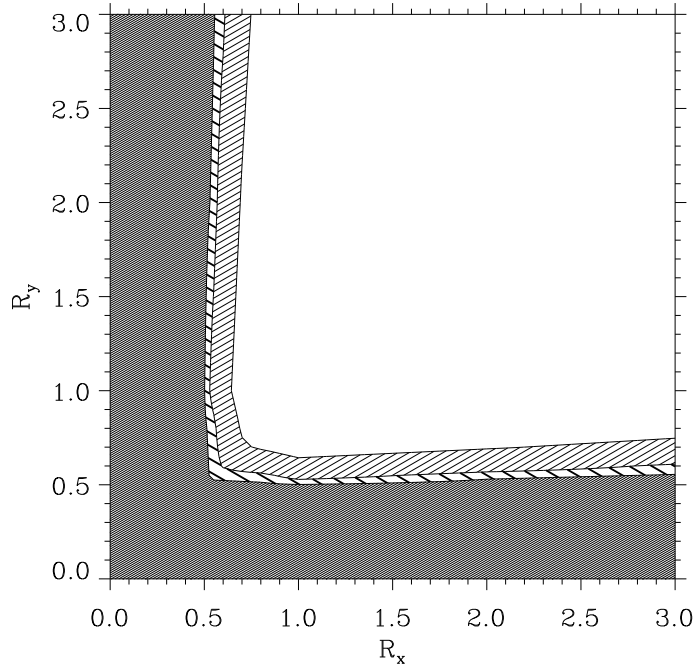}}}}
 \vspace{0.2cm}
 \caption{\small{
Grid of cumulative probability distributions of $S_{\circ}$ 
for $T^1$- and $T^2$-models obtained from Monte Carlo simulations. 
The thin-shaded, thick-shaded and grey regions correspond,
respectively, to the models ruled out at 68\%, 90\% and 95\% 
confidence.
}} 
 \label{fig:total}
 \end{figure}

A more complete picture of the cell size limits is obtained when we construct a
two-dimensional grid for different values of the cell sizes ($R_x,R_y,R_z$) 
with $R_z=3.0$ and $0.2<R_x,R_y<3.0$ (see Figure~\ref{fig:total}). 
The thin-shaded, thick-shaded and grey regions correspond, respectively, to 
the models ruled out at 68\%, 90\% and 95\%  confidence.
Notice in this plot that all contours are almost $L$-shaped, which means that, 
to a good approximation, the level in which a model ($R_x,R_y$) is ruled out 
depends only on the {\it smallest} cell size, $R_{min} \equiv {\it min} 
\{ R_x,R_y \}$. We see that $R_{min}\simgt$0.5 at 95\% confidence.

\section{Conclusions}

\tab
We have shown that the $COBE$/DMR maps have the ability to test and 
rule out $T^1$ and $T^2$ topological models. 
We have presented a new statistic to study these anisotropic models which 
quantifies the ``smallness" of a sky map in a single number, $S_{\circ}$, 
which is independent of the cell orientation, is precisely sensitive
to the type of symmetries that small universes produce, is easy to
work with and is easy to interpret.

From the $COBE$/DMR data, we obtain a lower limit for $T^1$- and $T^2$-models 
of $R_x\simgt$0.5, which corresponds to a cell size with smallest dimension
of $L$=3000$h^{-1}$Mpc. This limit is at 95\% confidence and assumes $n$=1. 
Since the topology is interesting only if the cell size is considerably
smaller than the horizon, so that it can (at least in principle) be directly 
observed, these models lose most of their appeal. 
Since the cubic $T^3$ case has already been ruled out as an interesting 
cosmological model \cite{dOCS95}, and $T^1$- and $T^2$-models for small 
cell sizes are ruled out, this means that {\it all} toroidal models (cubes and
rectangles) are ruled out as interesting cosmological models.

\acknowledgements{
We would like to thank Jon Aymon and Al Kogut for the help with the $COBE$
library subroutines and Max Tegmark for many 
useful comments on the manuscript. }

\end{document}